\documentclass[UTF8]{article}
\usepackage{geometry}
\usepackage{rotating}
\usepackage{amsmath,amsfonts}
\usepackage{algorithmic}
\usepackage{algorithm}
\usepackage{changes}
\definechangesauthor[color=teal]{R1}
\definechangesauthor[color=orange]{R2}
\definechangesauthor[color=purple]{R3}
\definechangesauthor[color=blue]{R2R3}
\usepackage{array}
\usepackage[position=t,singlelinecheck=off]{subfig}
\usepackage{textcomp}
\usepackage{stfloats}
\usepackage{url}
\usepackage{verbatim}
\usepackage{graphicx}
\usepackage{cite}
\usepackage{multirow}
\usepackage{booktabs}
\usepackage{color}
\title{Characterizing Regional Importance in Cities with Human Mobility Motifs in Metro Networks}

\author{Shuyang~Shi\footnote{Email: mail@shishuyang.cn}, Ding~Lyu, Lin~Wang, Xiaofan~Wang, and Guanrong~Chen}

\begin{document}

\maketitle

\begin{abstract}
Uncovering higher-order spatiotemporal dependencies within human mobility networks offers valuable insights into the analysis of urban structures. In most existing studies, human mobility networks are typically constructed by aggregating all trips without distinguishing who takes which specific trip. Instead, we claim individual mobility motifs, higher-order structures generated by daily trips of people, as fundamental units of human mobility networks. In this paper, we propose two network construction frameworks at the level of mobility motifs in characterizing regional importance in cities. Firstly, we enhance the structural dependencies within mobility motifs and proceed to construct mobility networks based on the enhanced mobility motifs. Secondly, taking inspiration from PageRank, we speculate that people would allocate values of importance to destinations according to their trip intentions. A motif-wise network construction framework is proposed based on the established mechanism. Leveraging large-scale metro data across cities, we construct three types of human mobility networks and characterize the regional importance by node importance indicators. Our comparison results suggest that the motif-based mobility network outperforms the classic mobility network, thus highlighting the efficacy of the introduced human mobility motifs. Finally, we demonstrate that the performance in characterizing the regional importance is significantly improved by our motif-wise framework.
\end{abstract}


\section{Introduction}

A city encompasses regions with diverse economic development orientation, and the fluxes of people between regions further reflect the complexity of the city. It holds great significance to scientifically understand the complexity of cities by studying the microscopic individual movement and the macroscopic flow of people, as well as their interactions with various urban regions. A deeper comprehension of human mobility within cities is a critical step in developing well-founded urban management strategies and transportation planning. Network science provides a method for abstracting the complex connections within a city in the form of networks\cite{batty2008size, pollock2016policy, xia2019ranking, zhou2023predicting, gu2018structuring, cheng2024spatial}. The regional importance, as reflected in human mobility network, serves as a vital tool for the study of city structure, which enables us to deeply understand the city structure\cite{liu2015revealing, tang2016statistical}, explore development and transformation process of city\cite{zhong2014detecting}, capture the city's circadian rhythm and assess the urban jobs-housing balance\cite{shi2022uncovering}. 


With the rapid development and proliferation of information technology, an increasing amount of social signal data related to human mobility is being collected, which paves the way for computing and studying the characteristics of both human mobility behavior and city structures\cite{barbosa2018human,zhou2018understanding,nilforoshan2023human, jia2023hierarchial}. 
In the scenario of smart cities, research on human mobility data, such as smart transportation card data, vehicle GPS data, and phone call logs, have been garnering more and more attention.
Metro networks serve as indispensable infrastructure for daily lives, work, and communications of urban residents. It indicates that the regional importance around metro stations can be characterized through the recognition and analysis of the diverse mobility patterns of passengers in metro networks. 

Previous studies typically define a human mobility network in metro systems as a directed weighted network, where nodes are metro stations and the weight of an edge denotes the magnitude of passenger flow between stations. In this conventional approach, the human mobility network is constructed by aggregating all trips without distinguishing who takes which path. For example, the aggregation by $A \rightarrow B$ of person 1 and $B \rightarrow C$ of person 2 is not distinguished from another by $A \rightarrow B \rightarrow C$ of one person. It has been pointed out that this kind of memoryless first-order Markov approach fails to capture the complete structure and dynamics of networks\cite{lambiotte2019networks,rosvall2014memory}, motivating the present exploration of higher-order dependencies in human mobility networks.

Existed studies on the higher-order organization of complex networks usually pre-define a variety of network motifs to explain structural mechanisms. This approach focuses on extracting or counting network motifs from a large-scale network to capture higher-order structural dependencies. However, in real networks, higher-order organizational features may not be solely governed by a single motif but rather emerge from the collective influence of multiple higher-order structures. In the context of human mobility networks, pre-defined network motifs may not comprehensively capture all mobility patterns. In \cite{schneider2013unravelling}, the human mobility motif is defined as a sequence of visited places and the trips among them, which can characterize various mobility patterns of passengers for the exploration of complex human behaviors. In this setting, an individual mobility motif can be presented as a mesoscopic network, in which nodes are the visited places and edges are the trips between visited places of a person in one day.

In recent years, a large number of achievements have been made in each of these areas individually. Relatively speaking, the exploration of the interrelationship of these fields remains a challenging topic. It is foreseeable that the complete trajectory of an individual's trip in a day can reflect more structure information of the human mobility network in the city  compared to multiple independent single trips. Building upon the above inspirations, our study of human mobility networks shifts from the aggregation of individuals' single trips to the aggregation of mobility motifs. Similarly to the approach in \cite{zhao2019ranking}, every single trip is considered as a first-order relation. A mobility motif consisting of all trips of a person between several stations in one day reflects higher-order network structures. Here, mobility motifs, also known as higher-order structures, are regarded as the fundamental units of a human mobility network. In this paper, we propose two novel network construction frameworks at the level of mobility motifs, and compare their effectiveness with the classic network construction method in characterizing regional importance through node importance indicators.

Firstly, we identify a variety of mobility motifs of people and then aggregate them to construct human mobility networks. To enhance the higher-order structural dependency, we reorganize mobility motifs with a strategy similar to those in \cite{benson2016higher,zhao2019ranking,li2021analyzing}. Then, we additively aggregate all reorganized mobility motifs to form a human mobility network, which is referred to as the motif-based mobility network. Secondly, we propose a new weight allocation method based on individual mobility motifs and trip intentions.  With the inspiration of PageRank, we reorganize mobility motifs by directly connecting the initial station and all destinations at different times and construct motif-wise network by aggregating these reorganized motifs. 

Specifically, the main contributions of this work are summarized as follows.
\begin{itemize}
  \item This work presents the first formulation and application of human mobility motifs to explore human mobility networks within cities and evaluate regional importance. We claim individual mobility motifs are the fundamental units of human mobility networks. It is a significant ideological transformation for the aggregation from first-order individuals' single trips to the higher-order mobility motifs in a human mobility network.
  \item We enhance the indirect effects to propose a motif-based aggregation framework for modeling human mobility networks. Instead of enumerating network motifs or subgraphs, this higher-order network construction framework uses mobility motifs as the fundamental units. The motif-based mobility network demonstrates superior performance when compared to the classic mobility network.
  \item We emphasize trip intentions by allocating station importance to destinations at different time periods in one day. Following the idea of PageRank, we propose a novel motif-wise framework, which can significantly improve the characterization of regional importance, as well as the model performance and explanation.
  \item We utilize transportation card data from three big cities in China to assess and compare the capabilities of the motif-based mobility network, motif-wise network, and classic mobility network in characterizing regional importance. House prices are used to evaluate the effectiveness of station rankings produced by our framework. Comparatively, we show that our motif-wise framework obtains the best performance.
\end{itemize}

\section{Related work}
\subsection{Higher-order organization of complex networks}
``Network motif'' was first defined by \cite{milo2002network} to characterize structural patterns of interconnections occurring in complex networks. Hidden mechanisms of higher-order structural dependencies in complex networks can be understood and explained by specific network motifs. In \cite{benson2016higher}, a variety of network motifs and the corresponding motif adjacency matrices are defined. A generic framework based on a selected motif adjacency matrix is developed to extract clusters organized with dense structural motifs. This motif-based framework has been extended and continuously improved to explore higher-order clustering structure\cite{yin2017local,xia2021chief}, community structure\cite{li2020community,gao2022higher},  core-periphery structure\cite{ma2018detection}. These findings help enhance the understanding of complex structural patterns and significantly broadens the application of complex network analysis\cite{chen2023composite, liu2023higher}.

Some classic work introduces node centrality indicators to assess the node importance by incorporating network motifs. In \cite{zhao2019ranking} and \cite{li2021analyzing}, induced networks from motif adjacency matrices are proposed and the PageRank or betweenness centrality is calculated for node importance ranking, with the proposed methods applied to construct motif-based cooperation networks for ranking scientists. Taking real h-indexes of scientists as the criterion, these models can improve the performance of ranking under specific motifs. However, little is known about how to select suitable motifs in different types of networks. Existing studies need to enumerate plenty of network motifs or simple subgraphs in order to obtain better higher-order expression. In addition, the higher-order connections of real-world networks may include multiple higher-order structures, rather than just one.

\subsection{Human mobility motifs}
Human mobility motifs were first proposed in \cite{schneider2013unravelling} to characterize diverse mobility patterns of people, which facilitated the exploration of human behaviors. The focus is on various mobility patterns hidden in the daily trips of people, revealing that most daily trips can be described by a few common mobility motifs. The daily mobility motifs are classified with four rules and a perturbation-based model is built to reproduce the generating of individual mobility motifs. Considering both location-based and activity-based motifs, in \cite{cao2019characterizing}, the scaling properties of travel distance are investigated, revealing the relationship between the scaling parameters and the node number of motifs. The findings indicate that people prefer a mobility motif with the lowest consumption that satisfies their demand, which is summarized as the least effort principle. In \cite{su2020pattern,su2021unveiling}, well-labeled human mobility motif data are used to present the differences between people in different working and living conditions in terms of the dimensions of motif complexity, travel time, travel purpose, etc. In \cite{liu2016using, jiang2017activity}, the major motif of each region in a city is analyzed, clarifying the relationship between the functions of these regions and the corresponding major motifs. A motif-preserving individual travel preference learning method is presented in \cite{chen2023multi}, and the accuracy and robustness of mobility prediction model is improved. Some classic research conclusions on individual movement patterns have also been extended at the level of location activity motifs, such as classical exploration and preferential return model\cite{xiong2021revealing}. Current research on human mobility motifs is devoted to exploring the variety of individual mobility patterns, but co-analyses with the global mobility network are rare.

\subsection{Human mobility patterns in metro networks and city structures}
Motivations for studying human mobility patterns in metro networks can be mainly divided into two categories: understanding human behaviors and analyzing network properties. Recognition and analysis of human mobility patterns provide a deep insight into complex human behaviors. Metro is a fundamental infrastructure associated with city structures, helping explore the latter specifically and precisely.

Several studies focus on passenger flow prediction and anomaly detection in metro systems. In \cite{ling2018predicting}, three machine learning methods are developed to detect anomalous passenger flows in metro networks. In \cite{chen2019subway}, an algorithm is proposed for predicting short-term irregular passenger flows during some special events, such as concerts or football matches. 
Taking both regional features and trip intentions into account, a human mobility prediction model is developed in \cite{shi2020prediction} with higher accuracy. When the function of a region changes, the model can predict fluxes in the new scene in accordance with the change. In \cite{he2015congestion}, the effects of two travel strategies are compared, respectively with the shortest path and minimum cost on the congestion in a metro network, where an intervention model is built, which shows that the congestion would obtain great mitigation as long as a small number of people switch travel strategies from the shortest path to the minimum cost.

Complex networks provide a powerful tool for analyzing city structures obtained from mobility data in metro networks. Network indicators and community structures have been widely used to describe the characteristics of human mobility and city structures\cite{xing2016weighted,xia2018exploring}. In \cite{zhong2014detecting}, the smart card data of Singapore in three years, from 2010 to 2012, are used to introduce structural indicators that can evaluate node centrality or detect communities for capturing changes in city structures. It was found that the distributions of the node centrality are becoming more and more flat, indicating that the urban development is becoming more and more balanced. In \cite{xia2019ranking}, a linear combination of node centrality, clustering coefficient, and closeness centrality is embedded into a basic PageRank framework for ranking the station importance. In \cite{shi2022uncovering}, directed weighted human mobility networks are built based on the three motif patterns identified through non-negative tensor decomposition. These networks, each constituted by one of the three motif patterns, exhibit distinct characteristics, shedding light on how different city regions serve diverse populations.

\section{Data description and processing}

\begin{figure}[!htb]
    \centering
    \includegraphics[width=\textwidth]{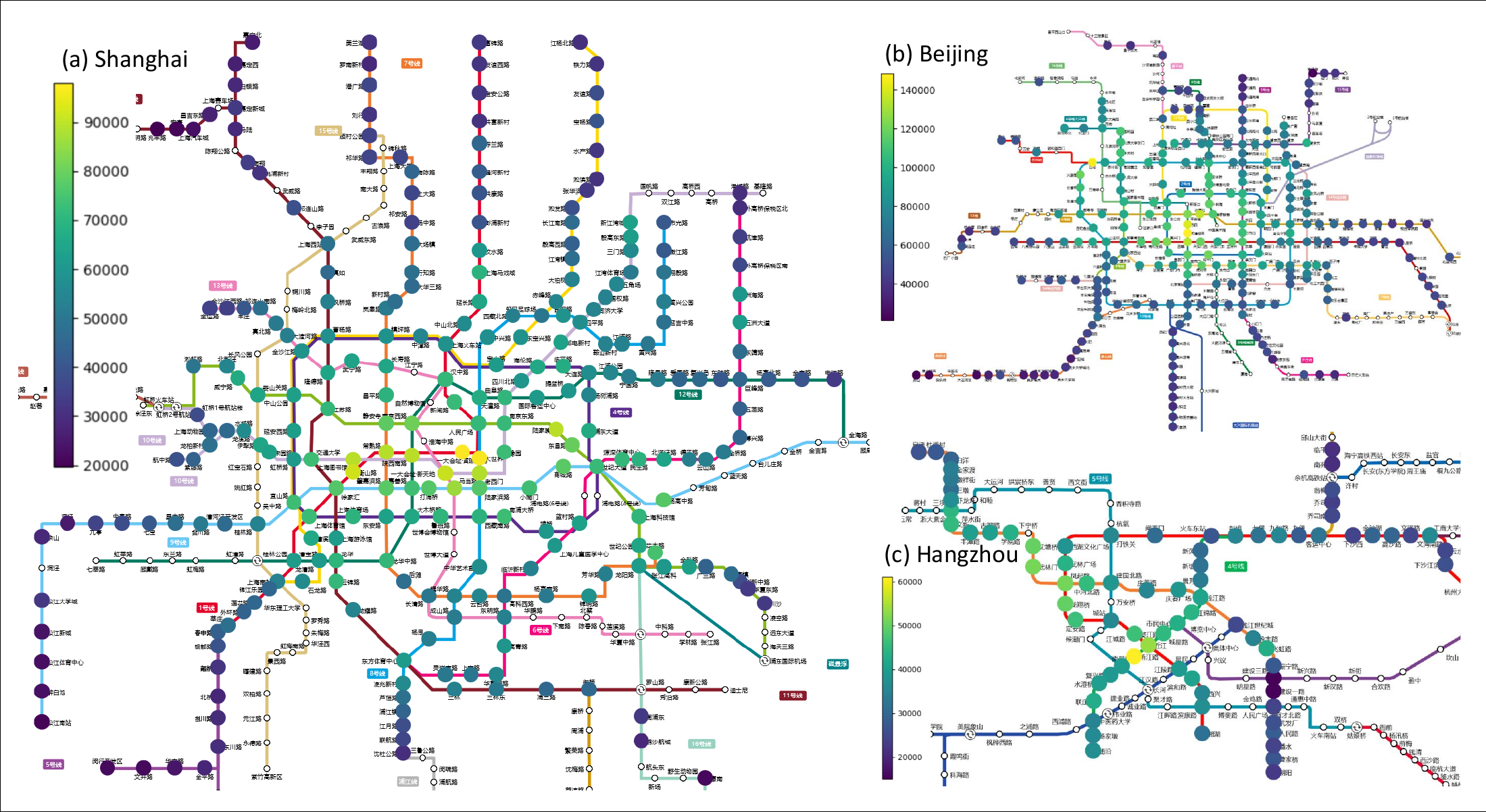}
    \caption{Topological structures and surrounding house prices of (a) Shanghai, (b) Beijing, and (c) Hangzhou Metro networks. Each node represents a metro station. The node color reflects the weighted average house price around a metro station, and the color bar shows the corresponding price in RMB. Newly built stations that haven't been color-filled are beyond the scope of this study. Few areas around metro stations (such as airports and industrial parks) do not have residential transaction records, which will be considered in the calculations of node importance but excluded in final ranking comparisons.}
    \label{fig:topology}
\end{figure}

We employ public transportation smart card data from Shanghai, Beijing, and Hangzhou metro in this study. Shanghai and Beijing are the largest cities with highly developed metro networks in China. The magnitude of daily metro passenger flows in Beijing and Shanghai was about 5 million during the studied period. To improve the robustness of our findings in this study, we also use Hangzhou metro data on a smaller network scale to verify possible variations. Topological structures of the metro networks in these three cities are visualized in Fig.~\ref{fig:topology}, and the basic information of Metro data is listed in Table~\ref{tab:data_summary}. Previous studies have pointed out that people's travel patterns may different on weekdays from weekends\cite{yong2018uncovering,xiong2021revealing}. Therefore, we separate the raw data of Shanghai and Hangzhou by weekdays and weekends for study.

\begin{table*}[!htb]
\centering
\resizebox{0.8\columnwidth}{!}{
\begin{tabular}{|l|l|l|l|l|}
\hline
Study City & Stations & Lines & Study Date & Motifs \\ \hline
Shanghai (weekdays) & 289 & 14 & 5-9 Sep 2016 & 11374267 \\ \hline
Shanghai (weekends) & 289 & 14 & 3,4,10,11,24,25 Sep 2016 & 13547178 \\ \hline
Beijing & 311 & 22 & 1 Mar 2018 & 2943270 \\ \hline
Hangzhou (weekdays) & 79 & 3 & 7-11,14-18  Jan 2019 & 6412542 \\ \hline
Hangzhou (weekends) & 79 & 3 & 5,6,12,13,19,20 Jan 2019 & 3349697 \\ \hline
\end{tabular}}
\caption{Summary of metro data.}\label{tab:data_summary}
\end{table*}

In this work, we aim to characterize regional importance in cities by analyzing passengers' mobility patterns in the metro networks. We speculate that regional importance can be characterized by the importance of metro stations in the consideration of human mobility. We expect to build human mobility networks in metro systems to predict the importance of surrounding areas of metro stations. As the saying goes, ``There are three factors that determine the house price: location, location, and location.'' In first-tier cities in China, house prices are a concentrated manifestation of the level of importance of regions. The stronger the importance of a place, the more attractive it is for people to live in, thereby driving up house prices. Therefore, house price is used in the evaluation of our frameworks. We crawl second-hand house prices around the metro stations during the studied years from a large real estate agency website\footnote{https://fang.com/}. The weighted average prices of houses around each metro station (total transaction price divided by total transaction area) are considered as the criterion to evaluate regional importance, which is shown in Fig.~\ref{fig:topology}. As commonly experienced, house prices are indeed higher in the downtown area and lower in the suburbs. It is also consistent with our intuitive cognition of regional importance in these cities.

\begin{figure}[!htb]
    \centering
    \includegraphics[width=\columnwidth]{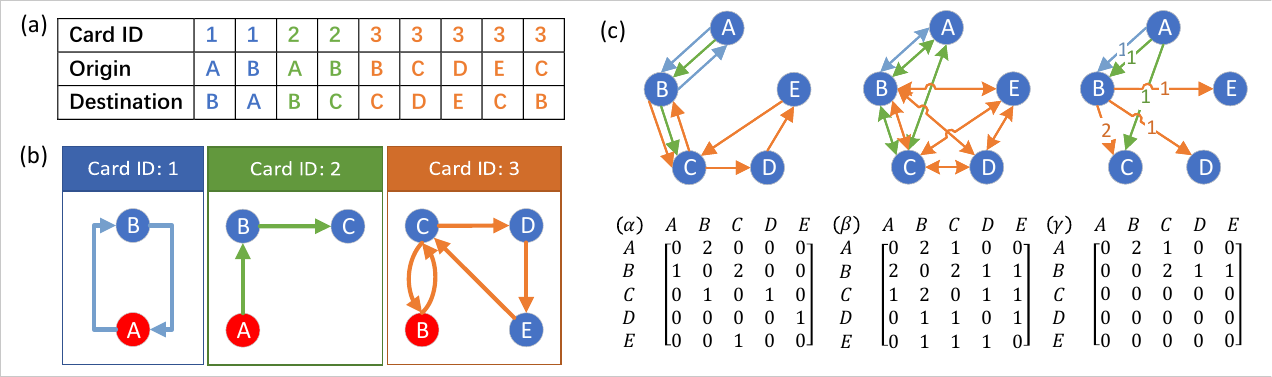}
    \caption{Illustration of three strategies of human mobility network construction. Consider a metro system with 5 stations (A-E) and 3 passengers (1-3). (a) Table list of nine Origin-Destination (OD) pairs extracted from raw transportation card data of three passengers in one day. Records with the same card ID have been sorted by time. Different colors are used to distinguish passengers and their trips. (b) Visualization of three mobility motifs generated by OD pairs. To clarify the direction of mobility motifs, the initial station of travel paths is marked in red. (c) Topological structures and corresponding adjacency matrices of human mobility networks with three different constructing strategies from mobility motifs: ($\alpha$) classic mobility network, ($\beta$) motif-based mobility network, and ($\gamma$) motif-wise mobility network. The edge color represents the passenger who contributes to the edge.}
    \label{fig:main_explain}
\end{figure}

\begin{figure}[!htb]
    \centering
    \includegraphics[width=0.7\columnwidth]{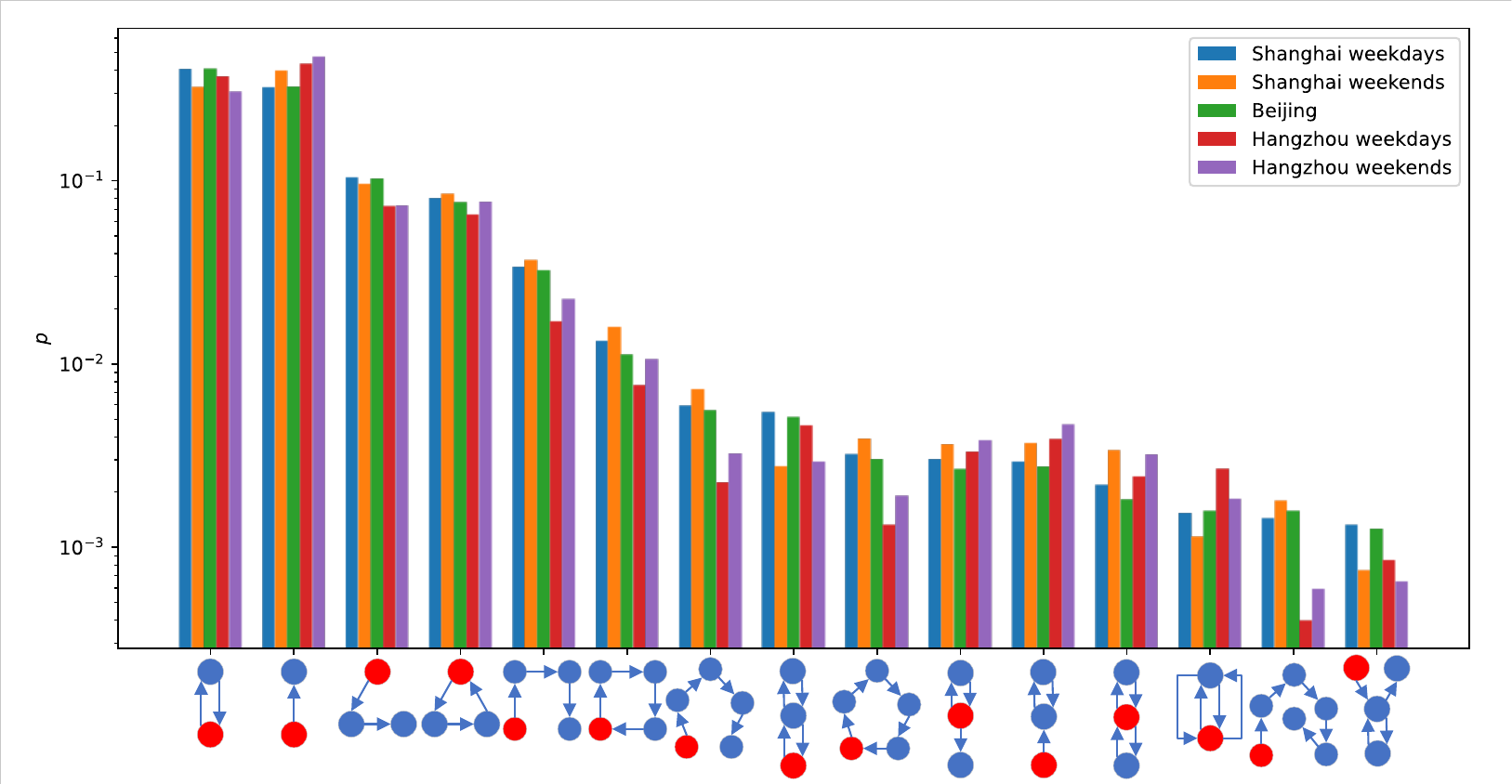}
    \caption{Distributions of the 15 most common daily human mobility motifs in Shanghai, Beijing, and Hangzhou. The statistics of mobility motifs on weekdays and weekends are separated.}
    \label{fig:moitf_distrubution}
\end{figure}

A human mobility motif is defined as an induced network from a passenger's daily trips in metro networks, in which nodes are the visited stations and edges are the trips between visited stations. From the raw transportation card data, we extract the daily OD pairs of each card. With the general assumption that each passenger is endowed with a unique Card ID, the trips of individuals are recorded as a set of temporal OD pairs. Taking stations as nodes and the trips between stations as directed edges, we obtain a variety of daily mobility motifs of individuals. Fig.~\ref{fig:main_explain}(a, b) illustrates the motifs generation process. Passenger 1 (with Card 1) first takes the metro from station $A$ to station $B$, and his mobility motif accordingly has a directed edge from $A$ to $B$. After that, this passenger returns to $A$ from $B$, so an edge from $B$ to $A$ is added to his mobility motif. The generation of two other mobility motifs is also shown in detail in Fig.~\ref{fig:main_explain}(a, b). The distributions of the 15 most common daily human mobility motifs in Shanghai, Beijing, and Hangzhou (in the descending order of the distributions according to Shanghai weekdays data) are presented in Fig.~\ref{fig:moitf_distrubution}. Moreover, the total number of all mobility motifs in the studied cities is counted in Table~\ref{tab:data_summary}.

\section{Methodology}
\subsection{Classic mobility networks}
Classic human mobility networks in the metro are generally defined as a directed weighted network $G=(V, W)$, where $V$ denotes the set of metro stations and $W=\{W_{ij}|i, j \in V, i\neq j\}$ denotes the set of the magnitudes of passenger flows between stations. As such, the network is the additive aggregation of all first-order trips. Meanwhile, human mobility networks can also be regarded as the aggregation of daily mobility motifs of all individuals.

\textbf{Definition 1} Let $M$ be a human mobility motif. The classic mobility network $G=(V, W)$ is defined as the additive aggregation of all such $M$s.

Although the networks constructed by Definition 1 and the general definition are the same, these two aggregation guidelines are hugely different. We no longer take the first-order trips (single OD pairs) as the fundamental units of a mobility network but take the higher-order mobility motifs as the fundamental units of a mobility network instead. We emphasize the idea of the higher-order organization of human mobility networks on the condition of mobility motifs as fundamental units, as visualized in Fig.~\ref{fig:main_explain}(c-$\alpha$). We aggregate three mobility motifs from Fig.~\ref{fig:main_explain}(b) other than directly aggregating single OD pairs from Fig.~\ref{fig:main_explain}(a) to construct the human mobility network.

\subsection{Motif-based mobility networks}
Previous studies on the higher-order organization of complex networks with network motifs introduced motif-based adjacency matrices\cite{benson2016higher,zhao2019ranking,li2021analyzing}. According to the definition, each element of a motif-based adjacency matrix equals the number of a specific motif that two nodes share. Formally, the motif-based network is defined as $\mathcal{G}=(V, \mathcal{W})$, where $\mathcal{W}_{ij}$ is the number of a specific motif containing both node $i$ and node $j$. From the meso perspective, each network motif is reorganized as a complete graph in which each pairwise node is connected by the motif. Thus, the motif-based mobility network is an additive aggregation of all complete graphs induced from network motifs.

Inspired by these studies, we adopt the same reorganization strategy on human mobility motifs. We generate mobility motifs from OD pair records of each passenger and reorganize them as fully connected graphs of the same nodes for the aggregation of a human mobility network.

\textbf{Definition 2} Let $M$ be a human mobility motif, and $\mathcal{M}$ be the induced undirected complete graph of $M$. The motif-based mobility network $\mathcal{G}=(V, \mathcal{W})$ is defined as the additive aggregation of all such $\mathcal{M}$s.

Fig.~\ref{fig:main_explain}(c-$\beta$) shows the construction of $\mathcal{G}$ by aggregating all reorganized motifs $\mathcal{M}$ in Fig.~\ref{fig:main_explain}(b). Compared with the classic mobility network, some new connections in the motif-based mobility network are established to enhance the latent effects between nodes. For example, there is a potential relevance between station $A$ and station $C$ in the mobility motif $M$ of person 2, so we bridge node $A$ and node $C$ with a directly connected edge when reorganizing the motif. Therefore, there is an edge between node $A$ and node $C$ in Fig.~\ref{fig:main_explain}(c-$\beta$) but not in Fig.~\ref{fig:main_explain}(c-$\alpha$). This new connection produced by $\mathcal{M}$ of person 2 enhances the inside higher-order structural dependencies due to the transform from indirect relevance to direct relevance. As a supplement, the motif-based mobility network is an undirected weighted network, with an adjacency matrix shown in Fig.~\ref{fig:main_explain}(c-$\beta$).

We will see later that the motif-based mobility networks have slightly better performance in characterizing regional importance than the classic mobility network. However, the accuracy of characterization still has room for improvement. With the inspiration of resource allocation, we further propose a novel aggregation framework to address the defects in the motif-based framework.

\subsection{Motif-wise human mobility networks}
\begin{figure}[!htb]
    \centering
    \includegraphics[width=0.7\columnwidth]{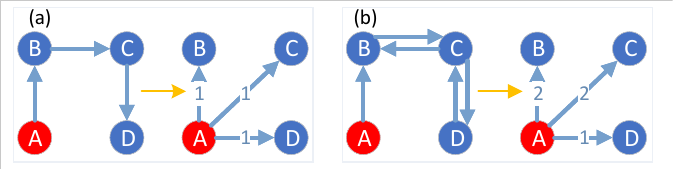}
    \caption{The reorganization strategy from human mobility motifs $M$ to $\mathbb{M}$.}
    \label{fig:motif_induced}
\end{figure}

According to the main idea of PageRank, each node is endued with an initial stochastic value of importance and then iteratively allocates the value into its out-neighbors until convergent. When considering mobility motifs of individuals, the value of nodes would be transferred along directed edges. However, some cases are out of our expectations. As an example, the significance of reorganization on a four-node mobility motif is visualized in Fig.~\ref{fig:motif_induced}(a). The person has a trip path $A \rightarrow B \rightarrow C \rightarrow D$, thus the mobility motif has four nodes and three directed edges. In such a case, the value of node $B$ will be taken out and allocated to node $C$ in the mobility motif, and the value of node $C$ is then allocated to node $D$. However, nodes $B$, $C$, and $D$ are all his destinations during different time periods, not just roles of transits. We expect that nodes $B$ and $C$ have similar importance with respect to node $D$. The contradiction to our expectations and the limitations of structural properties motivate us to ponder if there is a reasonable reorganization strategy on mobility motifs to address this problem. In this section, we propose a motif-wise framework to reorganize diverse mobility motifs of individuals and aggregate them to construct a human mobility network.

Inspired by the idea of PageRank, we adopt a new strategy to reorganize directed edges and redefine the edge weight. We speculate that a passenger would take his resources from his initial station to his travel destinations, instead of taking the resources from one destination to another. Consequently, assume that a passenger travels from the initial station with a value of resource (essentially the same as node importance). Once arriving at another location by metro, a corresponding value of resources is accordingly allocated to the destination station from the initial station.

\textbf{Definition 3} Let $V_M$ be the node set of a human mobility motif $M$, $s(i)$ be the weighted in-degree of node $i \in V_M$, and $A \in V_M$ be the initial node of $M$ . The reorganized motif of $M$ is defined as an induced graph $\mathbb{M}=(V_M, \mathbb{W})$, where $\mathbb{W}_{Ai}=s(i), i \in V_M\backslash A$.

As illustrated by Fig.~\ref{fig:motif_induced}(a), a passenger starts his trip from station $A$, and travels to $B$, $C$ and $D$ each once sequentially. Thus, the passenger allocates resources to these stations equally with the weight of 1. In Fig.~\ref{fig:motif_induced}(b), another passenger also travels to $B$, $C$ and $D$ from initial station $A$. While this passenger travels to $B$ and $C$ twice, so the directed edge weights $w_{AB}=w_{AC}=2$. Meanwhile, we also try another similar weight allocation method. The total resource of each person is restricted to be 1, equivalent to the sum of edge weights in a reorganized motif is 1. Each person allocates the value of resources to each place in proportion to the visiting frequency except the initial place. This normalizes the $\mathbb{W}_{Ai}$ in Definition 3 as $\mathbb{W}_{Ai}=s(i)/\sum_{j\in V_M\backslash A}s(j)$. For example, with this strategy, in Fig.~\ref{fig:motif_induced}(a), $w_{AB}=w_{AC}=w_{AD}=1/3$; and in Fig.~\ref{fig:motif_induced}(b), $w_{AB}=w_{AC}=2/5$, $w_{AD}=1/5$. We attach a mark to the former strategy with $w=1$ and the latter with $w=1/n$.

\textbf{Definition 4.} The motif-wise network $\mathbb{G}=(V, \mathbb{W})$ is defined as the additive aggregation of all such $\mathbb{M}s$.

Fig.~\ref{fig:main_explain}(c-$\gamma$) shows the construction process of the motif-wise mobility network based on the proposed framework with $w=1$. Passenger 1 only goes to station $B$ after starting from the initial station $A$, so he produces an edge with a weight of 1 from $A$ to $B$. Passenger 2 goes to station $B$ and $C$ once each after starting from the initial station $A$, so he produces an edge from $A$ to $B$ and an edge from $A$ to $C$ with weight of 1, respectively. Passenger 3 goes to $C$ twice, so the weight of the edge from initial station $B$ to $C$ is 2. He goes to $D$ and $E$ only once, so he produces an edge from $B$ to $D$ and an edge from $B$ to $E$ with weight of 1, respectively. We then aggregate all the reorganized motifs as a human mobility network and finally run node importance indicator calculation. Distinguished from the full connection of motif-based mobility networks, we enhance the directionality of mobility motifs.

\subsection{Measures of node importance}
In the proposed framework, one important task is to characterize the regional importance by ranking nodes in human mobility networks. Node importance can be calculated from four commonly used measures including PageRank value, eigenvector centrality, current flow closeness centrality, and clustering coefficient, which are respectively defined as follows.

\textbf{PageRank} The main idea of the weighted network PageRank is that each node assigns its PageRank value to the nodes it points to according to the edge weights. First, $\sum_{i=1}^{|V|} PR_i(0)=1$. Then, the PageRank values are calculated iteratively: 
$$PR_i(k)=s\sum_{j=1}^{|V|}\hat{w_{ji}}PR_j(k-1)+(1-s)\frac{1}{|V|},$$
where $\hat{w_{ji}}=w_{ji}/\sum_{k\in V} w_{jk}$, $|V|$ is the number of nodes and $s$ is a dampening factor, usually set as 0.85\cite{page1999pagerank, xia2019ranking}.

\textbf{Eigenvector centrality} The main idea of eigenvector centrality is that the importance of a node depends not only on the number of the neighboring nodes, but also on their importance. Let $x$ be the eigenvector of the largest eigenvalue of the adjacency matrix $A$. Then, the eigenvector centrality of node $i$ is the $i$-th element of $x$.

\textbf{Current flow closeness centrality}
Current-flow closeness centrality, also referred to as information centrality, calculates closeness centrality based on effective resistance between nodes in a network. When $G$ is regarded as an electrical network, the weight of each edge is equivalent to the conductance. The current flow closeness centrality of node $i$ is defined by
$$c_i=\frac{|V|-1}{\sum_{i\neq j}{p_{st}(i)-p_{st}(j)}},$$
where $p_{st}(i)-p_{st}(j)$ corresponds to the effective resistance.

\textbf{Clustering coefficient}
All regions of cities develop in coordination. Due to the existence of scale effects\cite{bettencourt2007growth,bettencourt2013origins} and network effects\cite{raimbault2020indirect}, the contact strength between neighboring regions of a region may also affect the development and importance of the region itself. Here, the clustering coefficient is used to capture the degree to which regions in a city tend to cluster together and treated as a node importance index. The clustering coefficient of node $i$ in a weighted network is defined as
$$c_i=\frac{1}{k(i)(k(i)-1)max(w)}\sum_{jk}(w_{ij}w_{ik}w_{jk})^{1/3},$$
where $k(i)$ is the degree of node $i$.

The larger the calculated values of these indicators, the higher the importance of the node.

\section{Results}
In this section, through real data analysis, we explain the improvement obtained by our frameworks and discuss some intriguing phenomena observed.
\subsection{Motif-base human mobility networks}
In this subsection, we calculate and rank the node importance in the two kinds of human mobility networks that we constructed above ($G$ and $\mathcal{G}$). For equivalent comparison with the undirected motif-based mobility network $\mathcal{G}$, we also convert $G$ to an undirected network $G'$. The edge weight of $G'$ is adjusted by $W'_{ij}=W'_{ji}=W_{ij}+W_{ji}$. As will be discussed later in Section \ref{section:motif-wise-discussion}, since $W_{ij}\approx W_{ji}$, the effect of the adjustment on the network structure is extremely small. The case of the directed network $G$ will be discussed in the next subsection.

Now we compare the station rankings calculated by the node importance measures with the weighted average house price around metro stations to evaluate the effectiveness of the ranking results. Following \cite{zhao2019ranking, li2021analyzing}, Normalized Discounted Cumulative Gain ($NDCG$) is used to evaluate the effectiveness of the ranking results, which is defined as
$$NDCG_k=\frac{DCG_k}{IDCG_k},$$
where $DCG_k=\sum_{i=1}^k \frac{rel_i}{\log_2(i+1)}$ is the discounted cumulative gain of the top $k$ stations obtained by a node importance indicator; house price is regarded as the relevance score $rel_i$ in our context; $IDCG_k$ is the $DCG$ of actual top $k$ stations with the highest house prices. This indicator ranges from 0 to 1. The larger the $NDCG$ is, the better the ranking performance is. We also introduce the Pearson correlation coefficient $r$ to depict the correlations.

\begin{figure}[!htb]
    \centering
    \includegraphics[width=\columnwidth]{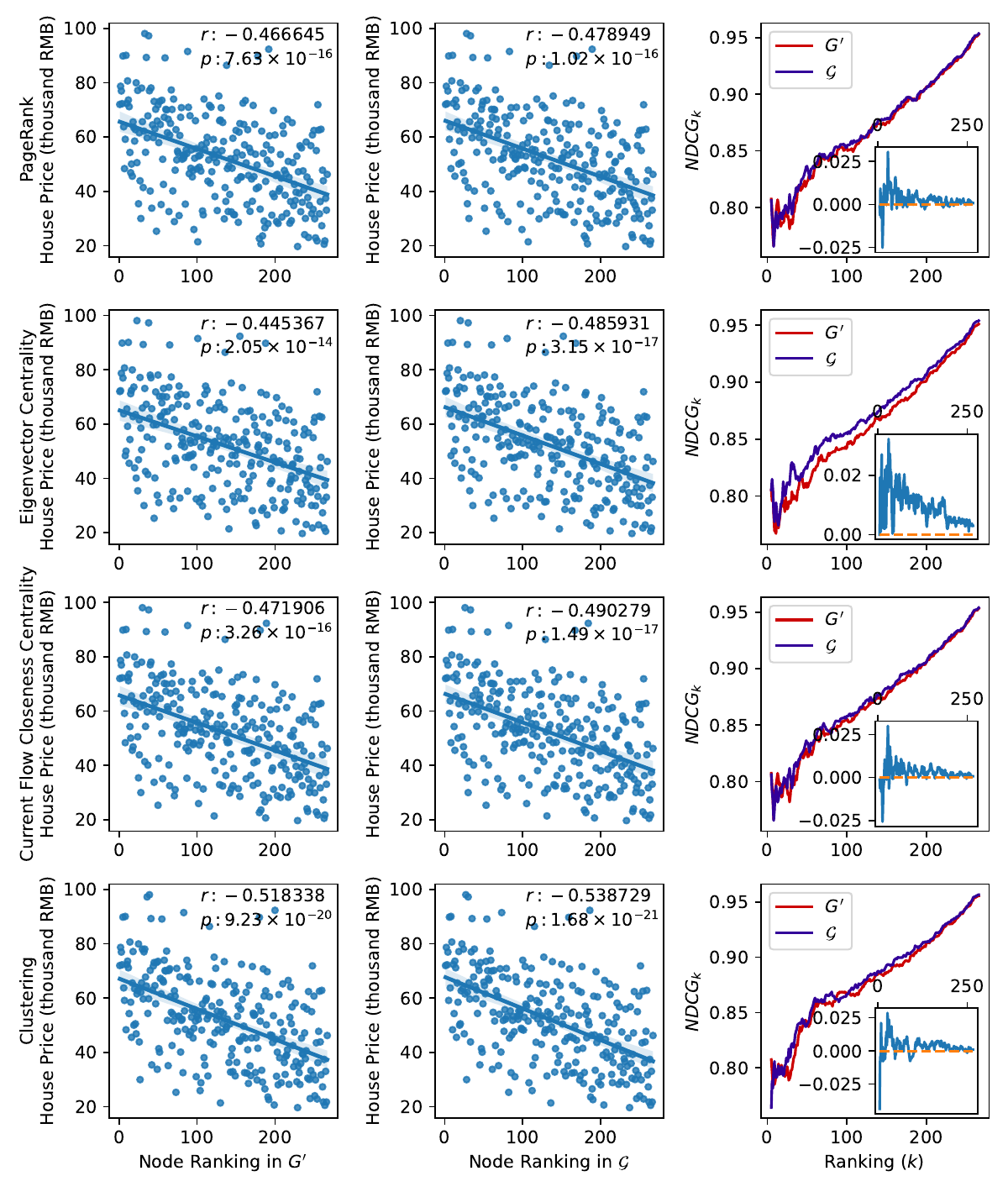}
    \caption{Correlation between Shanghai house price and calculated station rankings in $G$ and $\mathcal{G}$. Different rows show the ranking results by PageRank, eigenvector centrality, current flow closeness centrality, and clustering coefficient, respectively. The left two columns respectively depict the correlations of the weighted average house price against a measure of the rankings of metro stations in $G'$ and $\mathcal{G}$; 95\% confidence intervals are displayed. The Pearson correlation coefficient is shown in each subgraph. The right column compares $NDCG_k$ of $G'$ and $\mathcal{G}$. In order to present a clearer comparison, the inset window shows $NDCG_k$ of $\mathcal{G}$ minus $NDCG_k$ of $G'$. The orange dotted line is the datum line with $y=0$.}
    \label{fig:g1_main}
\end{figure}

We present a comparison of the data from Shanghai (weekdays) in Fig.~\ref{fig:g1_main} as an example and also list data from other cities to verify the robustness of the results in Table~\ref{tab:g1_main}. We take the ranking of node importance as the x-axis and house price as the y-axis in the first two columns in Fig.~\ref{fig:g1_main}. The trends of the negative correlations show that higher-ranking places have higher house prices, which is consistent with life experience. From the metrics of four node importance indicators, rankings of stations in $\mathcal{G}$ exhibit a slightly stronger linear correlation with house prices, as indicated by Pearson correlation coefficient $r$. Comparisons between the calculated $NDCG_k$ of $G'$ and $\mathcal{G}$ are shown in the right panel of Fig.~\ref{fig:g1_main}. To make it clearer, we make a difference calculation between $NDCG_k$ of $G'$ and $NDCG_k$ of $\mathcal{G}$ in the inset. $NDCG_k$ of $\mathcal{G}$ are larger than those of $G'$ with respect to most $k$s, indicating that motif-based mobility networks can help better characterize the regional importance. Furthermore, Table~\ref{tab:g1_main} shows the comprehensive results of the five studied objects according to the four metrics of node importance calculation. We highlight better results in bold. It can be seen that in all cities, the performance of motif-based mobility networks is generally better than that of classic mobility networks.

\begin{sidewaystable}[!htbp]
\centering
\begin{tabular}{|c|l|ll|ll|ll|ll|ll|}
\hline
\multicolumn{1}{|l|}{} &  & \multicolumn{2}{c|}{\begin{tabular}[c]{@{}c@{}}Shanghai\\ (weekdays)\end{tabular}} & \multicolumn{2}{c|}{\begin{tabular}[c]{@{}c@{}}Shanghai\\ (weekends)\end{tabular}} & \multicolumn{2}{c|}{Beijing} & \multicolumn{2}{c|}{\begin{tabular}[c]{@{}c@{}}Hangzhou\\ (weekdays)\end{tabular}} & \multicolumn{2}{c|}{\begin{tabular}[c]{@{}c@{}}Hangzhou\\ (weekends)\end{tabular}} \\ \hline
\multicolumn{1}{|l|}{Measures} & Indicators & \multicolumn{1}{l|}{$G'$} & $\mathcal{G}$ & \multicolumn{1}{l|}{$G'$} & $\mathcal{G}$ & \multicolumn{1}{l|}{$G'$} & $\mathcal{G}$ & \multicolumn{1}{l|}{$G'$} & $\mathcal{G}$ & \multicolumn{1}{l|}{$G'$} & $\mathcal{G}$ \\ \hline
\multirow{5}{*}{PageRank} & $r$ & \multicolumn{1}{l|}{-0.46665} & \textbf{-0.47895} & \multicolumn{1}{l|}{-0.39862} & \textbf{-0.41291} & \multicolumn{1}{l|}{-0.30185} & \textbf{-0.32199} & \multicolumn{1}{l|}{-0.37409} & \textbf{-0.37876} & \multicolumn{1}{l|}{-0.28968} & \textbf{-0.30941} \\ \cline{2-12}
 & $NDCG_5$ & \multicolumn{1}{l|}{\textbf{0.80752}} & 0.80636 & \multicolumn{1}{l|}{\textbf{0.76455}} & 0.73755 & \multicolumn{1}{l|}{0.53385} & \textbf{0.54724} & \multicolumn{1}{l|}{0.70932} & 0.70932 & \multicolumn{1}{l|}{0.74616} & 0.74616 \\ \cline{2-12}
 & $NDCG_{10}$ & \multicolumn{1}{l|}{0.78606} & \textbf{0.7909} & \multicolumn{1}{l|}{0.76598} & \textbf{0.76615} & \multicolumn{1}{l|}{0.60653} & \textbf{0.62435} & \multicolumn{1}{l|}{\textbf{0.77669}} & 0.76278 & \multicolumn{1}{l|}{0.74751} & \textbf{0.74803} \\ \cline{2-12}
 & $NDCG_{30}$ & \multicolumn{1}{l|}{0.78925} & \textbf{0.80109} & \multicolumn{1}{l|}{0.77105} & \textbf{0.79075} & \multicolumn{1}{l|}{0.65223} & \textbf{0.65897} & \multicolumn{1}{l|}{\textbf{0.83598}} & 0.83119 & \multicolumn{1}{l|}{0.8114} & \textbf{0.82537} \\ \cline{2-12}
 & $NDCG_{\infty}$ & \multicolumn{1}{l|}{0.95405} & \textbf{0.95487} & \multicolumn{1}{l|}{0.94823} & \textbf{0.94994} & \multicolumn{1}{l|}{0.91469} & \textbf{0.9183} & \multicolumn{1}{l|}{\textbf{0.93111}} & 0.93085 & \multicolumn{1}{l|}{0.92612} & \textbf{0.92707} \\ \hline
\multirow{5}{*}{\begin{tabular}[c]{@{}c@{}}Eigenvector\\ centrality\end{tabular}} & $r$ & \multicolumn{1}{l|}{-0.44537} & \textbf{-0.48593} & \multicolumn{1}{l|}{-0.39148} & \textbf{-0.42817} & \multicolumn{1}{l|}{-0.27852} & \textbf{-0.30874} & \multicolumn{1}{l|}{-0.39567} & \textbf{-0.40914} & \multicolumn{1}{l|}{-0.32824} & \textbf{-0.35526} \\ \cline{2-12}
 & $NDCG_5$ & \multicolumn{1}{l|}{0.80537} & 0.80537 & \multicolumn{1}{l|}{0.76455} & \textbf{0.80467} & \multicolumn{1}{l|}{\textbf{0.56548}} & 0.56316 & \multicolumn{1}{l|}{0.72803} & 0.72803 & \multicolumn{1}{l|}{0.74616} & 0.74616 \\ \cline{2-12}
 & $NDCG_{10}$ & \multicolumn{1}{l|}{0.76883} & \textbf{0.79306} & \multicolumn{1}{l|}{0.78463} & \textbf{0.7943} & \multicolumn{1}{l|}{0.57262} & \textbf{0.62769} & \multicolumn{1}{l|}{0.75086} & \textbf{0.75131} & \multicolumn{1}{l|}{0.73821} & \textbf{0.74613} \\ \cline{2-12}
 & $NDCG_{30}$ & \multicolumn{1}{l|}{0.79576} & \textbf{0.82797} & \multicolumn{1}{l|}{0.80284} & \textbf{0.80998} & \multicolumn{1}{l|}{0.61835} & \textbf{0.64429} & \multicolumn{1}{l|}{0.82442} & \textbf{0.82767} & \multicolumn{1}{l|}{0.80787} & \textbf{0.82032} \\ \cline{2-12}
 & $NDCG_{\infty}$ & \multicolumn{1}{l|}{0.95253} & \textbf{0.9555} & \multicolumn{1}{l|}{0.94964} & \textbf{0.95375} & \multicolumn{1}{l|}{0.91034} & \textbf{0.91676} & \multicolumn{1}{l|}{0.9287} & \textbf{0.9291} & \multicolumn{1}{l|}{0.92667} & \textbf{0.92774} \\ \hline
\multirow{5}{*}{\begin{tabular}[c]{@{}c@{}}Current flow\\ closeness\\ centrality\end{tabular}} & $r$ & \multicolumn{1}{l|}{-0.47191} & \textbf{-0.49028} & \multicolumn{1}{l|}{-0.40531} & \textbf{-0.42364} & \multicolumn{1}{l|}{-0.30237} & \textbf{-0.32131} & \multicolumn{1}{l|}{-0.38624} & \textbf{-0.39283} & \multicolumn{1}{l|}{-0.31552} & \textbf{-0.3281} \\ \cline{2-12}
 & $NDCG_5$ & \multicolumn{1}{l|}{\textbf{0.80752}} & 0.80636 & \multicolumn{1}{l|}{0.76455} & 0.76455 & \multicolumn{1}{l|}{0.55359} & \textbf{0.57103} & \multicolumn{1}{l|}{0.70932} & 0.70932 & \multicolumn{1}{l|}{0.74616} & 0.74616 \\ \cline{2-12}
 & $NDCG_{10}$ & \multicolumn{1}{l|}{\textbf{0.79219}} & 0.7909 & \multicolumn{1}{l|}{0.76598} & \textbf{0.76757} & \multicolumn{1}{l|}{0.62058} & \textbf{0.62564} & \multicolumn{1}{l|}{\textbf{0.77632}} & 0.75592 & \multicolumn{1}{l|}{0.74803} & 0.74803 \\ \cline{2-12}
 & $NDCG_{30}$ & \multicolumn{1}{l|}{\textbf{0.79853}} & 0.79825 & \multicolumn{1}{l|}{0.79281} & \textbf{0.79623} & \multicolumn{1}{l|}{0.65868} & \textbf{0.65887} & \multicolumn{1}{l|}{\textbf{0.83509}} & 0.83094 & \multicolumn{1}{l|}{0.83161} & \textbf{0.83389} \\ \cline{2-12}
 & $NDCG_{\infty}$ & \multicolumn{1}{l|}{0.95449} & \textbf{0.95525} & \multicolumn{1}{l|}{0.94871} & \textbf{0.95077} & \multicolumn{1}{l|}{0.91672} & \textbf{0.91816} & \multicolumn{1}{l|}{0.93091} & \textbf{0.93102} & \multicolumn{1}{l|}{0.92737} & \textbf{0.92755} \\ \hline
\multirow{5}{*}{Clustering} & $r$ & \multicolumn{1}{l|}{-0.51834} & \textbf{-0.53873} & \multicolumn{1}{l|}{-0.44969} & \textbf{-0.46788} & \multicolumn{1}{l|}{-0.34343} & \textbf{-0.36488} & \multicolumn{1}{l|}{-0.43742} & \textbf{-0.44209} & \multicolumn{1}{l|}{-0.36639} & \textbf{-0.38343} \\ \cline{2-12}
 & $NDCG_5$ & \multicolumn{1}{l|}{\textbf{0.80752}} & 0.76396 & \multicolumn{1}{l|}{0.73755} & \textbf{0.75747} & \multicolumn{1}{l|}{\textbf{0.57692}} & 0.56709 & \multicolumn{1}{l|}{0.70932} & 0.70932 & \multicolumn{1}{l|}{0.72815} & \textbf{0.74616} \\ \cline{2-12}
 & $NDCG_{10}$ & \multicolumn{1}{l|}{0.78795} & \textbf{0.79129} & \multicolumn{1}{l|}{0.76494} & \textbf{0.76554} & \multicolumn{1}{l|}{0.61572} & \textbf{0.62358} & \multicolumn{1}{l|}{\textbf{0.77699}} & 0.76303 & \multicolumn{1}{l|}{0.74765} & \textbf{0.74819} \\ \cline{2-12}
 & $NDCG_{30}$ & \multicolumn{1}{l|}{0.79776} & \textbf{0.82022} & \multicolumn{1}{l|}{0.79734} & \textbf{0.82093} & \multicolumn{1}{l|}{\textbf{0.67449}} & 0.66871 & \multicolumn{1}{l|}{0.83746} & \textbf{0.8377} & \multicolumn{1}{l|}{0.82552} & \textbf{0.83347} \\ \cline{2-12}
 & $NDCG_{\infty}$ & \multicolumn{1}{l|}{0.95701} & \textbf{0.95798} & \multicolumn{1}{l|}{0.95111} & \textbf{0.95279} & \multicolumn{1}{l|}{0.92056} & \textbf{0.92086} & \multicolumn{1}{l|}{0.93313} & \textbf{0.9333} & \multicolumn{1}{l|}{0.92919} & \textbf{0.92996} \\ \hline
\end{tabular}
\caption{Metric comparisons of the classic mobility network and the motif-based mobility network.}\label{tab:g1_main}
\end{sidewaystable}

\begin{figure}[!htb]
    \centering
    \includegraphics[width=0.9\columnwidth]{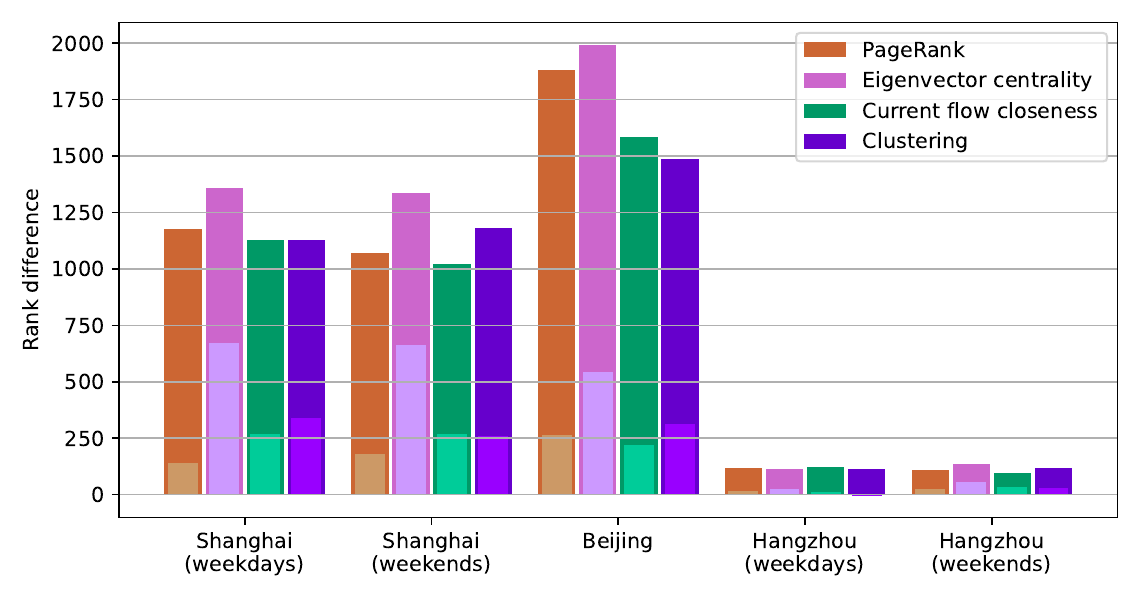}
    \caption{The upper bounds of the ranking improvements and the relative ranking improvements. The dark bars are the upper bounds and the light bars inside are those of the relative ranking improvements.}
    \label{fig:rank_diff_1}
\end{figure}

Next, we explain how much improvement is made by the motif-based mobility networks in characterizing regional importance ranking. Let $r_{G'}(i)$ and $r_\mathcal{G}(i)$ be the ranking of station $i$ calculated by a certain node importance indicator, $r_{hp}(i)$ be the house price ranking around station $i$. Then, $\sum_{i \in V}|r_{G'}(i)-r_\mathcal{G}(i)|$ is the sum of differences of the calculated rankings of the same station in the two mobility networks. This absolute difference of node rankings based on the two networks is the upper bound of ranking improvement. $\sum_{i \in V}(|r_{G'}(i)-r_{hp}(i)|-|r_\mathcal{G}(i)-r_{hp}(i)|)$ is the relative ranking improvement of $r_\mathcal{G}$ with respect to $r_{G'}$. Fig.~\ref{fig:rank_diff_1} shows the upper bounds of ranking improvements and the relative ranking improvements with different node importance indicators and different data sets. Except that the ranking improvement of Hangzhou weekdays by current flow closeness is $-2$, in all other cases, the motif-based mobility networks have made evident improvements in ranking compared with the classic mobility networks. Especially, when the eigenvector centrality is used as the indicator of node importance, the improvements are fairly obvious (this can also be observed from Fig.~\ref{fig:g1_main}).

From a network science perspective, we explore the higher-order structure of mobility networks by analyzing actual mobility motifs instead of exhaustively enumerating network motifs or simple subgraphs. This approach holds practical significance, and our results validate that employing higher-order network concepts can indeed yield better results in our scenario. This enhancement is not a result of selectively choosing favorable outcomes from an exhaustive search of network motifs or simple subgraphs.

\subsection{Motif-wise human mobility networks}
\begin{figure}[!htb]
    \centering
    \includegraphics[width=\hsize]{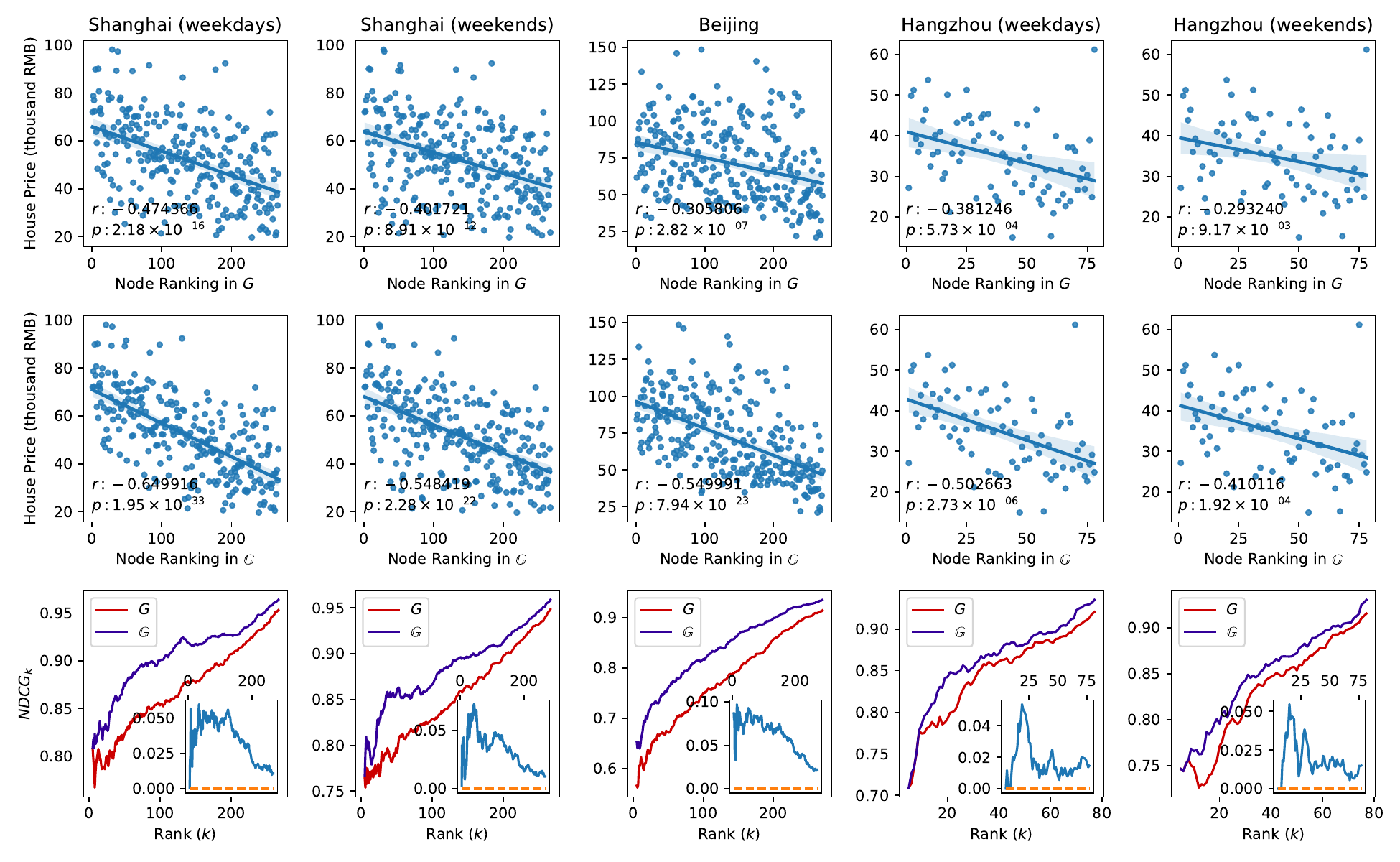}
    \caption{The correlation between house price and node ranking by PageRank. Data from different cities and time periods are divided into five columns. The top two rows respectively depict the correlations of the weighted average house price against a measure of the rankings of metro stations in $G$ and $\mathbb{G} (w=1)$; 95\% confidence intervals are displayed. The Pearson correlation coefficient is shown in each subgraph. $NDCG_k$ of $G$ and $\mathbb{G}$ are compared in the bottom row. }
    \label{fig:g2_main}
\end{figure}

It should be noted that people's travel patterns are directional. Therefore, we propose the motif-wise framework. We now compare the effectiveness of node importance ranking in motif-wise mobility networks $\mathbb{G}$ against classic mobility networks $G$ by PageRank. As shown in Fig.~\ref{fig:g2_main}, the accuracy of characterizing regional importance has increased significantly in all cities for all time periods.

Among the four node importance indicators mentioned above, eigenvector centrality can also be extended to directed graphs. We thus calculate the importance of nodes by eigenvector centrality and present the comparison results of our framework with the classic one. Meanwhile, as a comparison, we also establish 2-dimensional de Bruijn graph $G_{de}$ of human mobility\cite{lambiotte2019networks,rosvall2014memory}, where the memory nodes $\overrightarrow{ij}$ represent trips in metro and edges $\overrightarrow{ij} \rightarrow \overrightarrow{jk}$ represent connected trips, with weights $W(\overrightarrow{ij} \rightarrow \overrightarrow{jk})$ proportional to the passenger volume between stations and conditional on the previously visited station. The method can capture the second-order Markov process, which is another typical model useful for exploring the higher-order organization of complex systems.

We list comprehensive results of node importance ranking by PageRank and eigenvector centrality in human mobility networks in Tables.~\ref{tab:motif_wise_PageRank} and~\ref{tab:motif_wise_eigenvector}, respectively. In these two tables, we consider motif-wise mobility networks with the restriction of both $w=1$ and $w=1/n$. We highlight the better results in bold. Overall, it is obvious that $\mathbb{G}$ modeled by our framework has a significant improvement in node importance ranking than $G$ and $G_{de}$ in almost all aspects. From the perspective of node importance indicators, PageRank even performs slightly better than eigenvector centrality in characterizing regional importance, likely because our framework borrows the idea of PageRank. To our surprise, the results calculated by $G_{de}$ based on PageRank show poor performance. There is almost no correlation between the calculated node ranking and house price. The results on $NDCG$ are also far worse than all other methods. Only when using eigenvector centrality as the indicator, the correlation between node importance ranking and house price can be observed, where $G_{de}$ has a better performance compared to $G$ in Shanghai and Hangzhou. However, when $NDCG$ is used for assessment, $G_{de}$ still performs worse than $G$. It is probably because all motifs with only one trip in one day have to be ignored when constructing the de Bruijn graph but these motifs are non-negligible.

The proposed motif-wise framework has a performance boost compared with the motif-based framework according to $NDCG$. Moreover, the computed rankings of node importance display stronger linear correlations with house prices. The increasing performance is presented with comparisons between Table~\ref{tab:motif_wise_PageRank} (or Table~\ref{tab:motif_wise_eigenvector}) and Table~\ref{tab:g1_main}.

\begin{table*}[!htbp]
\centering
\begin{tabular}{|c|l|l|l|l|l|l|}
\hline
\multicolumn{1}{|l|}{} & Network & $r$ & $NDCG_{5}$ & $NDCG_{10}$ & $NDCG_{30}$ & $NDCG_{\infty}$ \\ \hline
\multirow{4}{*}{\begin{tabular}[c]{@{}c@{}}Shanghai\\ (weekdays)\end{tabular}} & $G$ & -0.47437 & 0.80752 & 0.79219 & 0.80156 & 0.95463 \\ \cline{2-7}
 & $G_{de}$ & 0.10465 & 0.58126 & 0.58793 & 0.61406 & 0.90752 \\ \cline{2-7}
 & $\mathbb{G} (w=1)$ & \textbf{-0.64992} & 0.80837 & 0.81415 & \textbf{0.84586} & \textbf{0.9654} \\ \cline{2-7}
 & $\mathbb{G} (w=1/n)$ & -0.63137 & \textbf{0.81346} & \textbf{0.81752} & 0.8363 & 0.96452 \\ \hline
\multirow{4}{*}{\begin{tabular}[c]{@{}c@{}}Shanghai\\ (weekends)\end{tabular}} & $G$ & -0.40172 & 0.76455 & 0.76598 & 0.78683 & 0.94858 \\ \cline{2-7}
 & $G_{de}$ & 0.00895 & 0.65538 & 0.6451 & 0.65953 & 0.91909 \\ \cline{2-7}
 & $\mathbb{G} (w=1)$ & \textbf{-0.54842} & \textbf{0.76689} & \textbf{0.795} & \textbf{0.83449} & \textbf{0.959} \\ \cline{2-7}
 & $\mathbb{G} (w=1/n)$ & -0.51458 & \textbf{0.76689} & 0.79202 & 0.81929 & 0.95598 \\ \hline
\multirow{4}{*}{Beijing} & $G$ & -0.30581 & 0.56548 & 0.60876 & 0.65616 & 0.91549 \\ \cline{2-7}
 & $G_{de}$ & 0.22892 & 0.46547 & 0.46941 & 0.50496 & 0.87686 \\ \cline{2-7}
 & $\mathbb{G} (w=1)$ & \textbf{-0.54999} & \textbf{0.65213} & \textbf{0.64326} & \textbf{0.73026} & \textbf{0.9366} \\ \cline{2-7}
 & $\mathbb{G} (w=1/n)$ & -0.5167 & 0.58032 & 0.63962 & 0.72527 & 0.93266 \\ \hline
\multirow{4}{*}{\begin{tabular}[c]{@{}c@{}}Hangzhou\\ (weekdays)\end{tabular}} & $G$ & -0.38125 & \textbf{0.70932} & 0.77438 & 0.837 & 0.93179 \\ \cline{2-7}
 & $G_{de}$ & 0.07817 & 0.51221 & 0.62363 & 0.72975 & 0.89179 \\ \cline{2-7}
 & $\mathbb{G} (w=1)$ & \textbf{-0.50266} & \textbf{0.70932} & \textbf{0.78535} & \textbf{0.85382} & \textbf{0.93772} \\ \cline{2-7}
 & $\mathbb{G} (w=1/n)$ & -0.49589 & 0.70421 & 0.78128 & 0.85182 & 0.93635 \\ \hline
\multirow{4}{*}{\begin{tabular}[c]{@{}c@{}}Hangzhou\\ (weekends)\end{tabular}} & $G$ & -0.29324 & \textbf{0.74616} & 0.74803 & 0.81195 & 0.92645 \\ \cline{2-7}
 & $G_{de}$ & 0.02113 & 0.67452 & 0.68211 & 0.75897 & 0.90957 \\ \cline{2-7}
 & $\mathbb{G} (w=1)$ & \textbf{-0.41012} & \textbf{0.74616} & \textbf{0.76545} & \textbf{0.84227} & \textbf{0.9328} \\ \cline{2-7}
 & $\mathbb{G} (w=1/n)$ & -0.39366 & 0.72139 & 0.75495 & 0.83956 & 0.93092 \\ \hline
\end{tabular}
\caption{PageRank ranking results of the classic mobility network and the motif-wise mobility network.}\label{tab:motif_wise_PageRank}
\end{table*}

\begin{table*}[!htbp]
\centering
\begin{tabular}{|c|l|l|l|l|l|l|}
\hline
\multicolumn{1}{|l|}{} & Network & $r$ & $NDCG_{5}$ & $NDCG_{10}$ & $NDCG_{30}$ & $NDCG_{\infty}$ \\ \hline
\multirow{4}{*}{\begin{tabular}[c]{@{}c@{}}Shanghai\\ (weekdays)\end{tabular}} & $G$ & -0.44611 & 0.81049 & 0.77046 & 0.79702 & 0.95287 \\ \cline{2-7}
 & $G_{de}$ & -0.44683 & 0.74952 & 0.71908 & 0.76939 & 0.94828 \\ \cline{2-7}
 & $\mathbb{G} (w=1)$ & \textbf{-0.63712} & 0.81784 & \textbf{0.81262} & \textbf{0.84221} & \textbf{0.96538} \\ \cline{2-7}
 & $\mathbb{G} (w=1/n)$ & -0.61425 & \textbf{0.81915} & 0.81244 & 0.8314 & 0.96378 \\ \hline
\multirow{4}{*}{\begin{tabular}[c]{@{}c@{}}Shanghai\\ (weekends)\end{tabular}} & $G$ & -0.38124 & \textbf{0.77995} & 0.78544 & 0.80291 & 0.94907 \\ \cline{2-7}
 & $G_{de}$ & -0.40532 & 0.72745 & 0.72244 & 0.76807 & 0.94611 \\ \cline{2-7}
 & $\mathbb{G} (w=1)$ & \textbf{-0.52934} & 0.76455 & \textbf{0.79377} & \textbf{0.8283} & \textbf{0.95801} \\ \cline{2-7}
 & $\mathbb{G} (w=1/n)$ & -0.49215 & 0.76675 & 0.79417 & 0.82718 & 0.95562 \\ \hline
\multirow{4}{*}{Beijing} & $G$ & -0.28222 & 0.56665 & 0.5684 & 0.61997 & 0.91154 \\ \cline{2-7}
 & $G_{de}$ & -0.20309 & 0.55528 & 0.55611 & 0.58689 & 0.90424 \\ \cline{2-7}
 & $\mathbb{G} (w=1)$ & \textbf{-0.51886} & \textbf{0.59871} & \textbf{0.63777} & \textbf{0.72506} & \textbf{0.93369} \\ \cline{2-7}
 & $\mathbb{G} (w=1/n)$ & -0.48905 & 0.56548 & 0.61076 & 0.70807 & 0.92774 \\ \hline
\multirow{4}{*}{\begin{tabular}[c]{@{}c@{}}Hangzhou\\ (weekdays)\end{tabular}} & $G$ & -0.40051 & \textbf{0.72803} & 0.75131 & 0.8254 & 0.9294 \\ \cline{2-7}
 & $G_{de}$ & -0.42487 & 0.71044 & 0.73556 & 0.82667 & 0.92903 \\ \cline{2-7}
 & $\mathbb{G} (w=1)$ & \textbf{-0.45648} & \textbf{0.72803} & \textbf{0.77936} & \textbf{0.83635} & \textbf{0.9334} \\ \cline{2-7}
 & $\mathbb{G} (w=1/n)$ & -0.4363 & 0.70421 & 0.77485 & 0.83165 & 0.93036 \\ \hline
\multirow{4}{*}{\begin{tabular}[c]{@{}c@{}}Hangzhou\\ (weekends)\end{tabular}} & $G$ & -0.32708 & 0.74616 & 0.75471 & 0.81699 & 0.92723 \\ \cline{2-7}
 & $G_{de}$ & \textbf{-0.40993} & 0.72916 & 0.75701 & 0.8362 & 0.93062 \\ \cline{2-7}
 & $\mathbb{G} (w=1)$ & -0.39722 & \textbf{0.7978} & \textbf{0.78793} & \textbf{0.84913} & \textbf{0.94149} \\ \cline{2-7}
 & $\mathbb{G} (w=1/n)$ & -0.38363 & 0.72139 & 0.75121 & 0.82658 & 0.92811 \\ \hline
\end{tabular}
\caption{Eigenvector centrality ranking results of the classic mobility network and motif-wise mobility network.}\label{tab:motif_wise_eigenvector}
\end{table*}

\begin{figure}[!htb]
    \centering
    \includegraphics[width=.8\columnwidth]{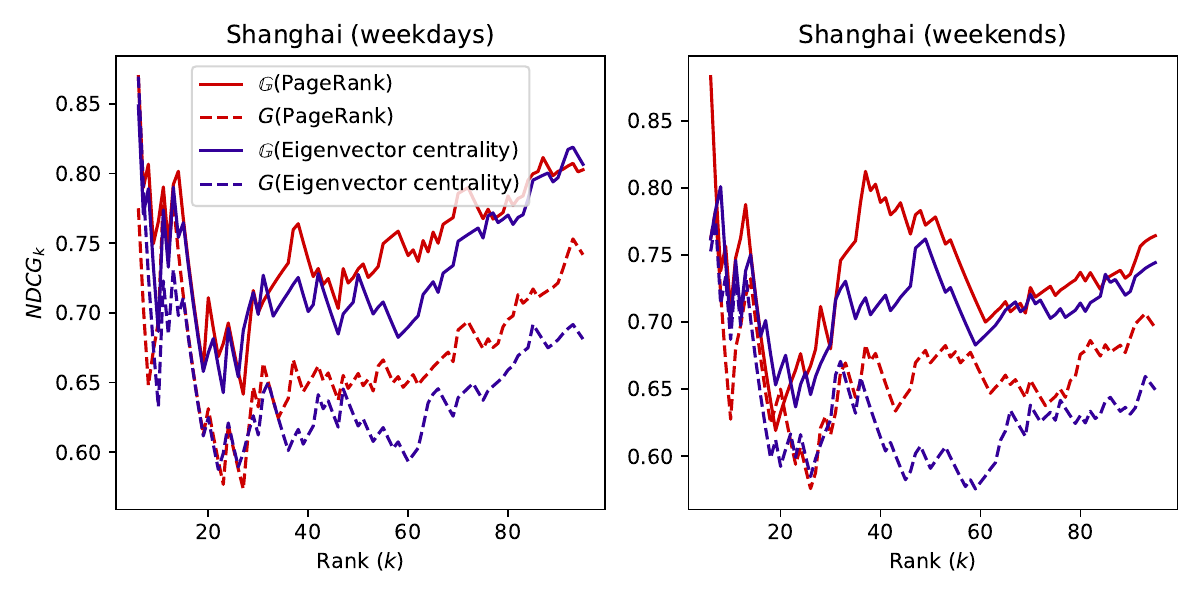}
    \caption{Comparisons between $G$ and $\mathbb{G}(w=1)$ critera on \textit{Shanghai Big Data Activity Report in 2017}.}
    \label{fig:yicai}
\end{figure}

To further illustrate the superiority of our framework, we evaluate the characterization results with an official report. Yicai, an official financial news arm of Shanghai Media Group\footnote{https://www.yicaiglobal.com/}, released the \textit{Shanghai Big Data Activity Report in 2017}. The report conducted a comprehensive analysis of the regional functions, consumer profiles, and business development in the vicinity of Shanghai metro stations. These factors serve as criteria for assessing regional importance. Top 100 metro stations are listed\footnote{https://metrocity.dtcj.com/shanghai}. Among these 100 stations, there are 5 newly built stations not shown in our data. We take the rankings of the remaining 95 stations as the criterion and use $NDCG$ again to compare the ranking results of our framework and the classic one. In this context, the relevance score $rel_i=1$ if the $i$-th important station calculated by the algorithm is also in the top-$k$ list of the criterion, otherwise $rel_i=0$. As shown in Fig.~\ref{fig:yicai}, $NDCG_k$ calculated by our framework (solid lines) are in the majority greater than those of traditional one (dashed lines), indicating that our framework again achieves significantly better performance.

\begin{figure}[!htb]
    \centering
    \includegraphics[width=0.9\columnwidth]{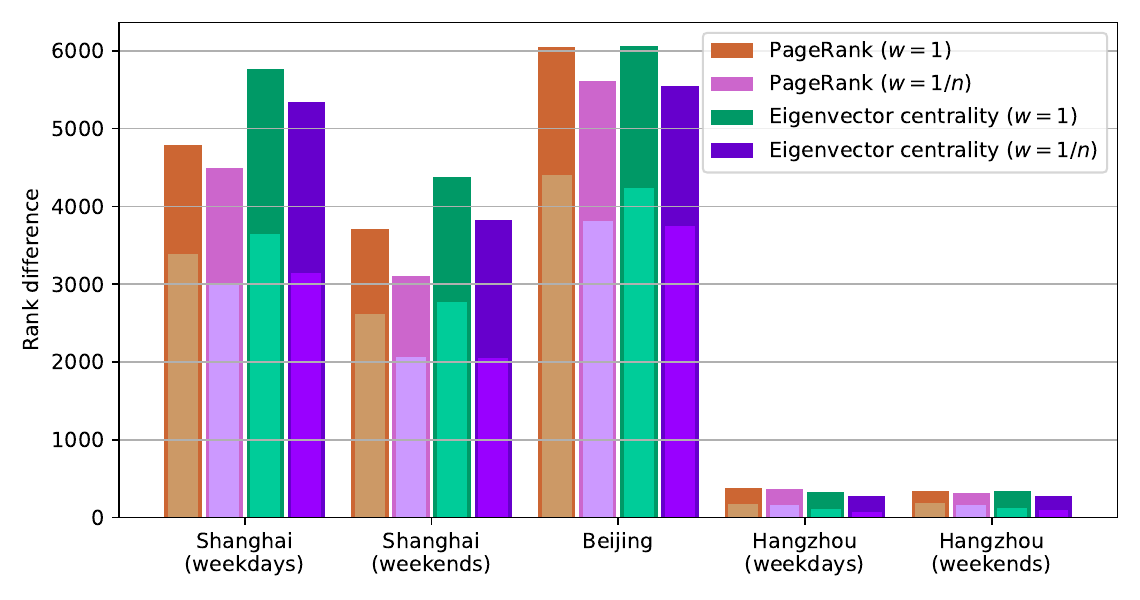}
    \caption{The upper bounds of the ranking improvements and the relative ranking improvements. The dark bars are the upper bounds and the light bars inside are those of the relative ranking improvements.}
    \label{fig:rank_diff_2}
\end{figure}

We show the upper bounds of the ranking improvements and the relative ranking improvements in Fig.~\ref{fig:rank_diff_2}. With our motif-wise framework, the upper bounds of the ranking improvements are over three times higher than that of the motif-based network. Moreover, the relative ranking improvements are much closer to the upper bounds of the ranking improvements in all cities. Overall, after the construction of the motif-wise mobility network, the node importance ranking performance is significantly improved.

\begin{figure}[!htb]
    \centering
    \includegraphics[width=0.8\columnwidth]{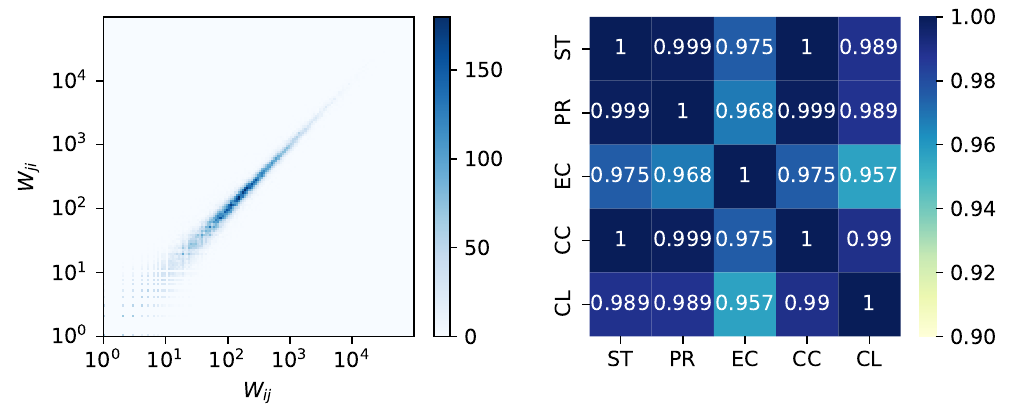}
    \caption{Relationship between in-strength and out-strength and the correlation matrix of node indicators of Shanghai (weekdays) data. (Left) The correlation between $W_{ij}$ and $W_{ji}$ of $G$. The colors indicate the numbers of the corresponding edges. (Right) Pearson correlation coefficients between node strength (the magnitude of passenger flows at stations) and four calculated node importance indicators in $G'$. ST: strength; PR: PageRank; EC: eigenvector centrality; CC: current flow closeness centrality; CL: clustering coefficient.}
    \label{fig:high_coff}
\end{figure}

\subsection{Discussion}\label{section:motif-wise-discussion}
We attempt to discuss the reasons for the differences in the results under the two network construction methods from the perspective of real urban scenarios. 

In China, there exist some areas in large cities called ``bedroom communities''. Many citizens live in these communities far from the city center just to sleep. Their working and entertainment places are probably far away from their living places, resulting in huge passenger flows at the metro stations around these communities. Since in human mobility network $G$, $W_{ij}\approx W_{ji}$, node importance is strongly associated with the strength of the node (the magnitude of passenger flow) according to the calculation of node importance indicators in the undirected weighted network (see Fig.~\ref{fig:high_coff}). However, the magnitude of passenger flows may not accurately represent the regional importance in the city. To further clarify this, we identify the top 10 stations with the largest declines in ranking obtained from the comparison of the classic framework and the proposed framework. We observe that these stations are primarily located in large residential communities in suburban areas. Taking Beijing as an example, as shown in Table~\ref{tab:Beijing_ranking}, the surrounding regions of such stations are mostly famous super-sized communities located in the suburbs. These areas experience super large passenger flows, leading to their high rankings in $G$. Yet, from a public perspective, these areas are not that comprehensively important. Our motif-wise framework addresses real human travel purposes and establishes a new directed weight allocation method, making the ranking results closer to reality by a large margin.

\begin{table*}[!htb]
\centering
\resizebox{.8\columnwidth}{!}{
\begin{tabular}{|l|l|l|l|}
\hline
Station Name & \begin{tabular}[c]{@{}l@{}}PageRank\\ranking in $G$\end{tabular} & \begin{tabular}[c]{@{}l@{}}PageRank\\ranking in $\mathbb{G}$\end{tabular} & \begin{tabular}[c]{@{}l@{}}Actual\\ ranking\end{tabular} \\ \hline
Changyang & 108 & 181 & 234 \\ \hline
Tiantongyuan North & 21 & 94 & 261 \\ \hline
Longze & 47 & 124 & 207 \\ \hline
Lishuiqiao & 35 & 115 & 176 \\ \hline
Wuzi Xueyuan Lu & 66 & 146 & 237 \\ \hline
Tiantongyuan & 13 & 96 & 256 \\ \hline
Jin'an Qiao & 44 & 128 & 228 \\ \hline
Huilong Guan & 41 & 136 & 220 \\ \hline
Huilong Guan Dongdajie & 59 & 179 & 227 \\ \hline
Huoying & 28 & 153 & 202 \\ \hline
\end{tabular}
}
\caption{The top 10 stations with the largest declines in ranking resulted from the two network construction methods.}\label{tab:Beijing_ranking}
\end{table*}

\section{Conclusion}
In the existing literature, significant strides have been made in the analysis of individual-level statistics and modeling, as well as in the exploration of population-level mobility patterns. Recent years have witnessed a growing emphasis on bridging the gap between meso-level individual travel patterns and macro-level population mobility, enriching our comprehension of the intricate dynamics of cities. This paper represents the inaugural endeavor to employ human mobility motifs in investigating urban mobility networks and assessing regional importance.

We first propose a bottom-up framework to model a motif-based mobility network that aggregates the reorganized individual mobility motifs by addition. Under this framework, the performance of characterizing the importance of regions is indeed improved. Comparing it to the previous top-down approach in exploring the higher-order organization of networks, our method eliminates the need for manual specification of motifs or subgraphs. The human mobility motifs that are aggregated to construct the whole network are all generated by real trips of passengers. This opens up a new direction for studying the higher-order properties of complex networks.

We then propose a weight allocation method by establishing a motif-wise framework, where individuals carry weights from their initial station to each destination station during trips. Under this framework, we aggregate the reorganized human mobility motifs to construct motif-wise networks. Compared with the classic mobility network, the performance of the motif-wise network in characterizing the importance of regions is greatly improved.

In summary, mobility motifs, also known as higher-order mobility structures, are considered as fundamental units of human mobility networks instead of first-order trips. Two motif reorganization methods are proposed under two frameworks, which enable the modeling of human mobility networks at the level of mobility motifs. The significance lies in that these cannot be achieved when only first-order single trips are considered. This underlines the potency of higher-order mobility motifs as a powerful tool for unraveling the higher-order organization within human mobility networks.

\bibliographystyle{IEEEtran}
\bibliography{main}

\begin{thebibliography}{10}
\providecommand{\url}[1]{#1}
\csname url@samestyle\endcsname
\providecommand{\newblock}{\relax}
\providecommand{\bibinfo}[2]{#2}
\providecommand{\BIBentrySTDinterwordspacing}{\spaceskip=0pt\relax}
\providecommand{\BIBentryALTinterwordstretchfactor}{4}
\providecommand{\BIBentryALTinterwordspacing}{\spaceskip=\fontdimen2\font plus
\BIBentryALTinterwordstretchfactor\fontdimen3\font minus \fontdimen4\font\relax}
\providecommand{\BIBforeignlanguage}[2]{{%
\expandafter\ifx\csname l@#1\endcsname\relax
\typeout{** WARNING: IEEEtran.bst: No hyphenation pattern has been}%
\typeout{** loaded for the language `#1'. Using the pattern for}%
\typeout{** the default language instead.}%
\else
\language=\csname l@#1\endcsname
\fi
#2}}
\providecommand{\BIBdecl}{\relax}
\BIBdecl

\bibitem{batty2008size}
M.~Batty, ``The size, scale, and shape of cities,'' \emph{science}, vol. 319, no. 5864, pp. 769--771, 2008.

\bibitem{pollock2016policy}
K.~Pollock, ``Policy: urban physics,'' \emph{Nature}, vol. 531, no. 7594, pp. S64--S66, 2016.

\bibitem{xia2019ranking}
F.~Xia, J.~Wang, X.~Kong, D.~Zhang, and Z.~Wang, ``Ranking station importance with human mobility patterns using subway network datasets,'' \emph{IEEE Transactions on Intelligent Transportation Systems}, vol.~21, no.~7, pp. 2840--2852, 2019.

\bibitem{zhou2023predicting}
Z.~Zhou, K.~Yang, Y.~Liang, B.~Wang, H.~Chen, and Y.~Wang, ``Predicting collective human mobility via countering spatiotemporal heterogeneity,'' \emph{IEEE Transactions on Mobile Computing}, 2023.

\bibitem{gu2018structuring}
T.~Gu, M.~Zhu, W.~Chen, Z.~Huang, R.~Maciejewski, and L.~Chang, ``Structuring mobility transition with an adaptive graph representation,'' \emph{IEEE Transactions on Computational Social Systems}, vol.~5, no.~4, pp. 1121--1132, 2018.

\bibitem{cheng2024spatial}
Y.~Cheng, C.~Li, Y.~Zhang, S.~He, and J.~Chen, ``Spatial–temporal urban mobility pattern analysis during covid-19 pandemic,'' \emph{IEEE Transactions on Computational Social Systems}, vol.~11, no.~1, pp. 38--50, 2024.

\bibitem{liu2015revealing}
X.~Liu, L.~Gong, Y.~Gong, and Y.~Liu, ``Revealing travel patterns and city structure with taxi trip data,'' \emph{Journal of transport Geography}, vol.~43, pp. 78--90, 2015.

\bibitem{tang2016statistical}
J.~Tang, S.~Zhang, W.~Zhang, F.~Liu, W.~Zhang, and Y.~Wang, ``Statistical properties of urban mobility from location-based travel networks,'' \emph{Physica A: Statistical Mechanics and its Applications}, vol. 461, pp. 694--707, 2016.

\bibitem{zhong2014detecting}
C.~Zhong, S.~M. Arisona, X.~Huang, M.~Batty, and G.~Schmitt, ``Detecting the dynamics of urban structure through spatial network analysis,'' \emph{International Journal of Geographical Information Science}, vol.~28, no.~11, pp. 2178--2199, 2014.

\bibitem{shi2022uncovering}
S.~Shi, L.~Wang, and X.~Wang, ``Uncovering the spatiotemporal motif patterns in urban mobility networks by non-negative tensor decomposition,'' \emph{Physica A: Statistical Mechanics and its Applications}, vol. 606, p. 128142, 2022.

\bibitem{barbosa2018human}
H.~Barbosa, M.~Barthelemy, G.~Ghoshal, C.~R. James, M.~Lenormand, T.~Louail, R.~Menezes, J.~J. Ramasco, F.~Simini, and M.~Tomasini, ``Human mobility: Models and applications,'' \emph{Physics Reports}, vol. 734, pp. 1--74, 2018.

\bibitem{zhou2018understanding}
Y.~Zhou, B.~P.~L. Lau, C.~Yuen, B.~Tun{\c{c}}er, and E.~Wilhelm, ``Understanding urban human mobility through crowdsensed data,'' \emph{IEEE Communications Magazine}, vol.~56, no.~11, pp. 52--59, 2018.

\bibitem{nilforoshan2023human}
H.~Nilforoshan, W.~Looi, E.~Pierson, B.~Villanueva, N.~Fishman, Y.~Chen, J.~Sholar, B.~Redbird, D.~Grusky, and J.~Leskovec, ``Human mobility networks reveal increased segregation in large cities,'' \emph{Nature}, pp. 1--7, 2023.

\bibitem{jia2023hierarchial}
W.~Jia, K.~Zhao, and S.~Zhao, ``The hierarchical clustering of human mobility behaviors,'' \emph{IEEE Transactions on Computational Social Systems}, pp. 1--12, 2023.

\bibitem{lambiotte2019networks}
R.~Lambiotte, M.~Rosvall, and I.~Scholtes, ``From networks to optimal higher-order models of complex systems,'' \emph{Nature Physics}, vol.~15, no.~4, pp. 313--320, 2019.

\bibitem{rosvall2014memory}
M.~Rosvall, A.~V. Esquivel, A.~Lancichinetti, J.~D. West, and R.~Lambiotte, ``Memory in network flows and its effects on spreading dynamics and community detection,'' \emph{Nature Communications}, vol.~5, no.~1, pp. 1--13, 2014.

\bibitem{schneider2013unravelling}
C.~M. Schneider, V.~Belik, T.~Couronn{\'e}, Z.~Smoreda, and M.~C. Gonz{\'a}lez, ``Unravelling daily human mobility motifs,'' \emph{Journal of The Royal Society Interface}, vol.~10, no.~84, p. 20130246, 2013.

\bibitem{zhao2019ranking}
H.~Zhao, X.~Xu, Y.~Song, D.~L. Lee, Z.~Chen, and H.~Gao, ``Ranking users in social networks with motif-based pagerank,'' \emph{IEEE Transactions on Knowledge and Data Engineering}, 2019.

\bibitem{benson2016higher}
A.~R. Benson, D.~F. Gleich, and J.~Leskovec, ``Higher-order organization of complex networks,'' \emph{Science}, vol. 353, no. 6295, pp. 163--166, 2016.

\bibitem{li2021analyzing}
X.~Li, R.~Cheng, K.~C.-C. Chang, C.~Shan, C.~Ma, and H.~Cao, ``On analyzing graphs with motif-paths,'' \emph{Proceedings of the VLDB Endowment}, vol.~14, no.~6, pp. 1111--1123, 2021.

\bibitem{milo2002network}
R.~Milo, S.~Shen-Orr, S.~Itzkovitz, N.~Kashtan, D.~Chklovskii, and U.~Alon, ``Network motifs: simple building blocks of complex networks,'' \emph{Science}, vol. 298, no. 5594, pp. 824--827, 2002.

\bibitem{yin2017local}
H.~Yin, A.~R. Benson, J.~Leskovec, and D.~F. Gleich, ``Local higher-order graph clustering,'' in \emph{Proceedings of the 23rd ACM SIGKDD International Conference on Knowledge Discovery and Data Mining}, 2017, pp. 555--564.

\bibitem{xia2021chief}
F.~Xia, S.~Yu, C.~Liu, J.~Li, and I.~Lee, ``Chief: Clustering with higher-order motifs in big networks,'' \emph{IEEE Transactions on Network Science and Engineering}, vol.~9, no.~3, pp. 990--1005, 2022.

\bibitem{li2020community}
P.-Z. Li, L.~Huang, C.-D. Wang, J.-H. Lai, and D.~Huang, ``Community detection by motif-aware label propagation,'' \emph{ACM Transactions on Knowledge Discovery from Data (TKDD)}, vol.~14, no.~2, pp. 1--19, 2020.

\bibitem{gao2022higher}
Y.~Gao, H.~Zhang, and X.~Yu, ``Higher-order community detection: On information degeneration and its elimination,'' \emph{IEEE/ACM Transactions on Networking}, 2022.

\bibitem{ma2018detection}
C.~Ma, B.-B. Xiang, H.-S. Chen, M.~Small, and H.-F. Zhang, ``Detection of core-periphery structure in networks based on 3-tuple motifs,'' \emph{Chaos: An Interdisciplinary Journal of Nonlinear Science}, vol.~28, no.~5, p. 053121, 2018.

\bibitem{chen2023composite}
J.~Chen, M.~Feng, D.~Zhao, C.~Xia, and Z.~Wang, ``Composite effective degree markov chain for epidemic dynamics on higher-order networks,'' \emph{IEEE Transactions on Systems, Man, and Cybernetics: Systems}, vol.~53, no.~12, pp. 7415--7426, 2023.

\bibitem{liu2023higher}
M.~Liu, J.~Li, Z.~Yang, and K.~Yang, ``Higher-order functional structure exploration in heterogeneous combat network based on operational motif spectral clustering,'' \emph{IEEE Systems Journal}, vol.~17, no.~3, pp. 4279--4290, 2023.

\bibitem{cao2019characterizing}
J.~Cao, Q.~Li, W.~Tu, and F.~Wang, ``Characterizing preferred motif choices and distance impacts,'' \emph{PloS One}, vol.~14, no.~4, p. e0215242, 2019.

\bibitem{su2020pattern}
R.~Su, E.~C. McBride, and K.~G. Goulias, ``Pattern recognition of daily activity patterns using human mobility motifs and sequence analysis,'' \emph{Transportation Research Part C: Emerging Technologies}, vol. 120, p. 102796, 2020.

\bibitem{su2021unveiling}
------, ``Unveiling daily activity pattern differences between telecommuters and commuters using human mobility motifs and sequence analysis,'' \emph{Transportation Research Part A: Policy and Practice}, vol. 147, pp. 106--132, 2021.

\bibitem{liu2016using}
Z.~Liu, J.~Yu, W.~Xiong, J.~Lu, J.~Yang, and Q.~Wang, ``Using mobile phone data to explore spatial-temporal evolution of home-based daily mobility patterns in shanghai,'' in \emph{2016 International Conference on Behavioral, Economic and Socio-cultural Computing (BESC)}.\hskip 1em plus 0.5em minus 0.4em\relax IEEE, 2016, pp. 1--6.

\bibitem{jiang2017activity}
S.~Jiang, J.~Ferreira, and M.~C. Gonzalez, ``Activity-based human mobility patterns inferred from mobile phone data: A case study of singapore,'' \emph{IEEE Transactions on Big Data}, vol.~3, no.~2, pp. 208--219, 2017.

\bibitem{chen2023multi}
Y.~Chen, N.~Xie, H.~Xu, X.~Chen, and D.-H. Lee, ``A multi-context aware human mobility prediction model based on motif-preserving travel preference learning,'' \emph{IEEE Transactions on Intelligent Transportation Systems}, pp. 1--14, 2023.

\bibitem{xiong2021revealing}
Q.~Xiong, Y.~Liu, P.~Xie, Y.~Wang, and Y.~Liu, ``Revealing correlation patterns of individual location activity motifs between workdays and day-offs using massive mobile phone data,'' \emph{Computers, Environment and Urban Systems}, vol.~89, p. 101682, 2021.

\bibitem{ling2018predicting}
X.~Ling, Z.~Huang, C.~Wang, F.~Zhang, and P.~Wang, ``Predicting subway passenger flows under different traffic conditions,'' \emph{PLoS One}, vol.~13, no.~8, p. e0202707, 2018.

\bibitem{chen2019subway}
E.~Chen, Z.~Ye, C.~Wang, and M.~Xu, ``Subway passenger flow prediction for special events using smart card data,'' \emph{IEEE Transactions on Intelligent Transportation Systems}, vol.~21, no.~3, pp. 1109--1120, 2019.

\bibitem{shi2020prediction}
S.~Shi, L.~Wang, S.~Xu, and X.~Wang, ``Prediction of intra-urban human mobility by integrating regional functions and trip intentions,'' \emph{IEEE Transactions on Knowledge and Data Engineering}, 2020.

\bibitem{he2015congestion}
K.~He, Z.~Xu, P.~Wang, L.~Deng, and L.~Tu, ``Congestion avoidance routing based on large-scale social signals,'' \emph{IEEE Transactions on Intelligent Transportation Systems}, vol.~17, no.~9, pp. 2613--2626, 2015.

\bibitem{xing2016weighted}
Y.~Xing, J.~Lu, and S.~Chen, ``Weighted complex network analysis of shanghai rail transit system,'' \emph{Discrete Dynamics in Nature and Society}, vol. 2016, 2016.

\bibitem{xia2018exploring}
F.~Xia, J.~Wang, X.~Kong, Z.~Wang, J.~Li, and C.~Liu, ``Exploring human mobility patterns in urban scenarios: A trajectory data perspective,'' \emph{IEEE Communications Magazine}, vol.~56, no.~3, pp. 142--149, 2018.

\bibitem{yong2018uncovering}
N.~Yong, S.~Ni, S.~Shen, P.~Chen, and X.~Ji, ``Uncovering stable and occasional human mobility patterns: A case study of the beijing subway,'' \emph{Physica A: Statistical Mechanics and its Applications}, vol. 492, pp. 28--38, 2018.

\bibitem{page1999pagerank}
L.~Page, S.~Brin, R.~Motwani, and T.~Winograd, ``The pagerank citation ranking: Bringing order to the web.'' Stanford InfoLab, Tech. Rep., 1999.

\bibitem{bettencourt2007growth}
L.~M. Bettencourt, J.~Lobo, D.~Helbing, C.~K{\"u}hnert, and G.~B. West, ``Growth, innovation, scaling, and the pace of life in cities,'' \emph{Proceedings of the National Academy of Sciences}, vol. 104, no.~17, pp. 7301--7306, 2007.

\bibitem{bettencourt2013origins}
L.~M. Bettencourt, ``The origins of scaling in cities,'' \emph{science}, vol. 340, no. 6139, pp. 1438--1441, 2013.

\bibitem{raimbault2020indirect}
J.~Raimbault, ``Indirect evidence of network effects in a system of cities,'' \emph{Environment and Planning B: Urban Analytics and City Science}, vol.~47, no.~1, pp. 138--155, 2020.

\end{thebibliography}
\end{document}